\documentclass[conference]{IEEEtran}
\pdfoutput=1

\usepackage{amsmath}  
\usepackage{amsfonts}  
\usepackage{graphicx}
\usepackage[sort]{cite}

\begin{document}

 \title{BER Performance of Uplink Massive {MIMO} \\  With Low-Resolution ADCs  }
 
 \author{\IEEEauthorblockN{Azad Azizzadeh}
 \IEEEauthorblockA{Department of Electrical Engineering\\
 Razi University, Kermanshah, Iran.\\
 Email: azizzadeh.azad@stu.razi.ac.ir}
 \and
 \IEEEauthorblockN{Reza Mohammadkhani}
 \IEEEauthorblockA{Department of Electrical Engineering\\
 University of Kurdistan, Sanandaj, Iran.\\
 Email: r.mohammadkhani@uok.ac.ir}
 
\and
\IEEEauthorblockN{Seyed Vahab Al-Din Makki}
\IEEEauthorblockA{Department of Electrical Engineering\\
Razi University, Kermanshah, Iran.\\
Email: v.makki@razi.ac.ir}}


\maketitle

\begin{abstract}
Massive multiple-input multiple-output (MIMO) is a promising technology for next generation wireless communication systems (5G). In this technology, Base Station (BS) is equipped with a large number of antennas. Employing high resolution analog-to-digital converters (ADCs) for all antennas may cause high costs and high power consumption for the BS. 

By performing numerical results, we evaluate the use of low-resolution ADCs for  uplink massive MIMO by analyzing Bit Error Rate (BER) performance for different detection techniques (MMSE, ZF) and different modulations (QPSK, 16-QAM) to find an optimal quantization resolution.
Our results reveal that the BER performance of uplink massive MIMO systems with a few-bit resolution ADCs is comparable to the case of having full precision ADCs. We found that the optimum choice of quantization level (number of bits in ADCs) depends on the modulation technique and the number of antennas at the BS.
\end{abstract}


\section{Introduction}
Massive MIMO technology as a result of rethinking the concept of MIMO wireless communications, enables each base station (BS) to communicate with tens of users at the same time and frequency, by increasing the number of antennas at the BS \cite{rusek2013scaling, larsson2014massive}. In these systems, each antenna is  followed by a radio frequency (RF) chain (including an ADC unit).
However,  power consumption and hardware complexity of ADCs grow exponentially by increasing the quantization resolution \cite{walden1999analog}. This makes a critical problem for massive MIMO technology having hundreds of antennas at the BS. 

 A solution to this problem is to employ low-resolution ADCs \cite{risi2014massive,wang2014multiuser,mo2015capacity}. However, these studies have investigated the ultimate coarse quantization of 1-bit ADCs. 
The symbol error rate (SER) of an uplink massive MU-MIMO system employing 1-bit ADCs and transmit modulation QPSK is analysed in \cite{risi2014massive,wang2014multiuser}, whereas \cite{mo2015capacity} evaluates the capacity for the aforesaid scenario.

As an alternative solution, a mixed-ADC architecture is suggested in \cite{zhang2016mixed,liang2016mixed}, where many one-bit ADCs are used along with a few high precision ADCs on some of the antennas. They found that adding a few number of high-resolution ADCs makes a significant enhancement of the BER performance.

Due to non-linear behaviour of the quanitzation function, theoretical analysis of the system performance, especially BER criterion, would be very difficult. Therefore, In this paper, we study the BER performance of uplink massive MIMO systems with different coarse quantization levels of $b$-bit resolution ADCs for two modulation schemes of QPSK and 16-QAM, through simulation results.

The rest of this paper is organized as follows. In Section \ref{sec:model}, we describe the system model and quantization method used in the numerical simulation.

\begin{figure}[t]
\centering
\includegraphics[width=0.99\linewidth]{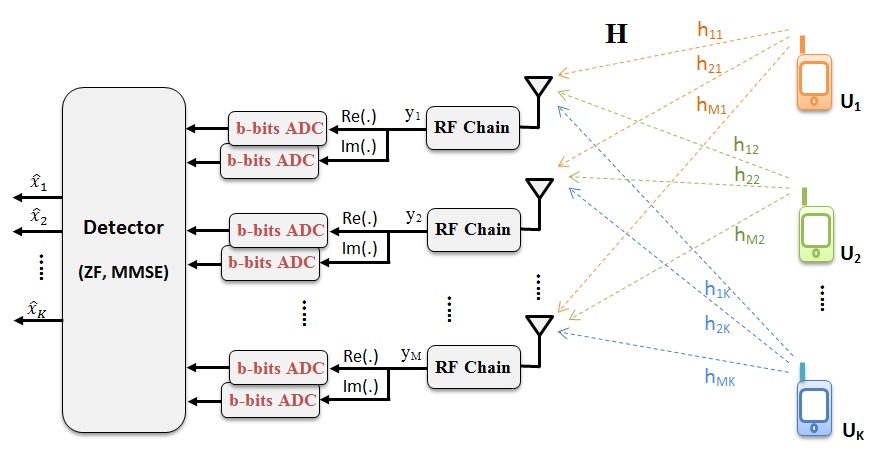}
\caption{Quantized massive MU-MIMO uplink system model}
\label{fig:ADC_massive}
\end{figure}

\section{System Model}\label{sec:model}
We consider a single-cell uplink multiuser MIMO system of one BS communicating with K users. We further assume the BS has $M$ antennas, while all users are considered to be single-antenna.  As shown in Fig.\ref{fig:ADC_massive}, the baseband received signal vector at the BS is given by
\begin{equation}
\mathbf y =\mathbf  H \mathbf x +\mathbf n=\sum_{k=1}^K \mathbf h_k x_k+\mathbf n
\end{equation}
where $\mathbf h_k \in \mathbb{C}^{M\times 1}$ is the channel vector between the BS and the $k$th user, $\mathbf H\overset{\Delta}{=} [\mathbf h_1,\mathbf h_2,...,\mathbf h_K]\in \mathbb{C}^{M\times K}$ is the channel matrix, $\mathbf x \in \mathbb{C}^{K\times 1}$  denotes the vector of the transmitted symbols from all users, $\mathbf y \in \mathbb{C}^{M\times 1}$ is the received signal vector before quantization, and $\mathbf n \in \mathbb{C}^{M\times 1}$  is the additive white Gaussian noise vector.

As illustrated in Fig.\ref{fig:ADC_massive}, quantized version of the complex received signal at each antenna, can be obtained by using two $b$-bit ADCs for both the real and imaginary parts.
 The resulting quantized signal vector is defined as
\begin{equation}
\mathbf r=Q(\mathbf y)=Q(\Re(\mathbf y))+j.Q(\Im(\mathbf y))
\end{equation}
where $Q(\cdot)$ represents the quantization function, and it is a non-linear function. 

In the simulation results, we assume that  $Q(\cdot)$ is a $b$-bit uniform mid-riser quantizer with $N=2^b$ levels, that converts the real input signal $y \in [y_i,y_{i+1})$ to a real-valued output $r_i$ for $i=1,2,\dots,N$ \cite{gersho1991vector}. The input interval endpoints $y_i$ are
\begin{eqnarray}
y_i =
\begin{cases} 
    -\infty            & i=1\\
     (-N/2-1+i) \Delta & i=2,3,...,N \\
     +\infty           & i=N+1
\end{cases} 
\end{eqnarray}
where $\Delta$ is the quantization step-size. The quantizer output values $r_i$ are defined as 
\begin{equation}
r_i = (-\frac{N}{2}-\frac{1}{2}+i) \Delta, \hspace{10pt}   i=1,2,...,N
\end{equation}

\begin{figure}
\centering
\includegraphics[width=0.9\linewidth]{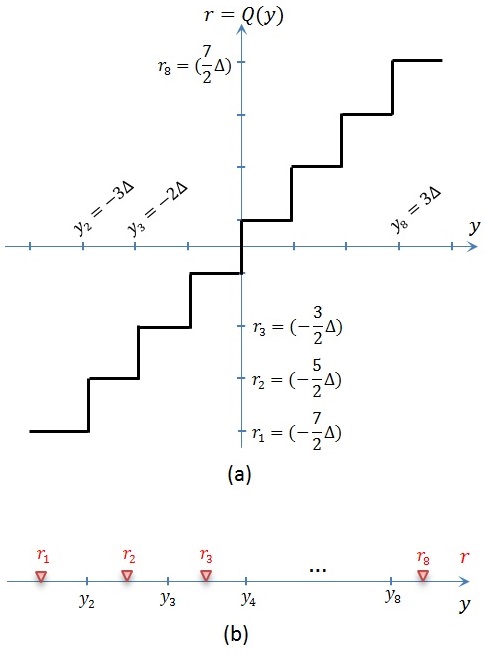}
\caption{(a) 2D staircase, and (b) 1D horizontal representations of the input-output characteristic of a 3-bit uniform mid-riser quantizer}
\label{fig:midrise_quantizer }
\end{figure}

Fig. \ref{fig:midrise_quantizer } shows the input-output characteristic of a 3-bit uniform mid-riser quantizer.

\section{Uplink MIMO Detection}
In general, a MIMO detector is used to estimate the transmitted symbol vector $\mathbf x$  from the received signal vector $\mathbf y$. Maximum likelihood (ML) detector provides the optimal performance. However its complexity grows dramatically by increasing the number of antennas and constellation order \cite{brown2012practical}.  By contrast, linear detectors have low computational complexity achieving suboptimal performance \cite{brown2012practical,ma2014massive}, such as zero forcing (ZF) and minimum mean square error (MMSE) detectors. For the case of massive MIMO where $M\gg K\gg 1$, it is shown that those linear detectors perform properly well \cite{bai2012low,ngo2013energy}. Therefore we employ ZF and MMSE detectors in this paper.

We assume that perfect channel state information (CSI) is available at the BS, in other words, it knows the channel matrix $\mathbf H$ perfectly. Applying a linear detector with an $M\times K$  detection matrix $\mathbf A$ to the received signal vector $\mathbf y$, an estimate of the transmit vector $\mathbf x$ can be expressed as
\begin{equation}
\hat{\mathbf x} =\mathbf A^H \mathbf y=\mathbf A^H(\mathbf H\mathbf x+\mathbf n)
\end{equation}
where $\hat{\mathbf x}=[\hat x_1,\hat x_2,...,\hat x_K]^T$ is a $K\times 1$ signal vector consists of the data streams from the $K$ single-antenna devices. Therefore, we have the $kth$ element of $\hat{\mathbf x}$ as:
\begin{equation}
\hat x_k =\underbrace{\mathbf a_k^H \mathbf h_k x_k }_\text{desired signal}+ \underbrace{\sum_{i\neq k}^K \mathbf a_k^H \mathbf h_i x_i}_\text{interuser interference}+\underbrace{\mathbf a_k^H \mathbf n}_\text{noise}
\end{equation}

In this paper, we employ ZF and MMSE linear detectors by defining  $\mathbf A$ as \cite{ma2014massive}
\begin{eqnarray}
\mathbf A=
\begin{cases} 
    \mathbf H (\mathbf H^H\mathbf H)^{-1}                                                                                                      &\text{for ZF}\\
   \mathbf H (\mathbf H^H\mathbf H+\frac{\sigma^2_n}{\sigma^2_x} \mathbf I_K)^{-1}                &\text{for MMSE}
\end{cases} 
\end{eqnarray}
where $\sigma^2_x$ and $\sigma^2_n$  are the transmit signal and the received noise variances respectively. As we see from above, the detection matrix $\mathbf A$ requires the knowledge of the channel matrix $\mathbf H$.

For a quantized MIMO system, we have
\begin{equation}
\hat{\mathbf x} =\mathbf A^H \mathbf r=\mathbf A^H\mathbf Q(\mathbf y)=\mathbf A^H\mathbf Q(\mathbf H\mathbf x+\mathbf n).
\end{equation}

However, due to the presence of non-linear function $Q(\cdot)$, an exact analysis of BER for quantized Massive MIMO system is rather difficult, while it is relatively simple to perform simulation studies. We use numerical results, to evaluate the BER performance of the quantized system, employing low resolution ADCs.

\section{Simulation Results}\label{sec:sim}
In this section, Monte Carlo simulations are used to demonstrate the effect of low resolution ADCs on the BER performance of uplink massive MIMO systems.  We consider the channel $\mathbf H$ to be i.i.d Rayleigh fading and the noise at each antenna as an additive zero-mean white complex Gaussian. Two transmission modulation schemes of QPSK and 16-QAM are employed by the $K$-users, and the BS uses ZF and MMSE detection techniques. Simulations are performed with $M = 100$ antennas at the BS  serving $K=10$ users. We consider different $b$-bit ADC resolutions by applying $b = 1, 2, 3, 4$ and $\infty$. We note that $b=\infty$ corresponds to the case of full precision quantization in the simulation results.
The BER Monte Carlo simulation values are obtained by averaging over 100 channel realizations, preceded by an averaging across all the users for each channel realization.

\begin{figure}[t]
\centering
\includegraphics[width=0.99\linewidth]{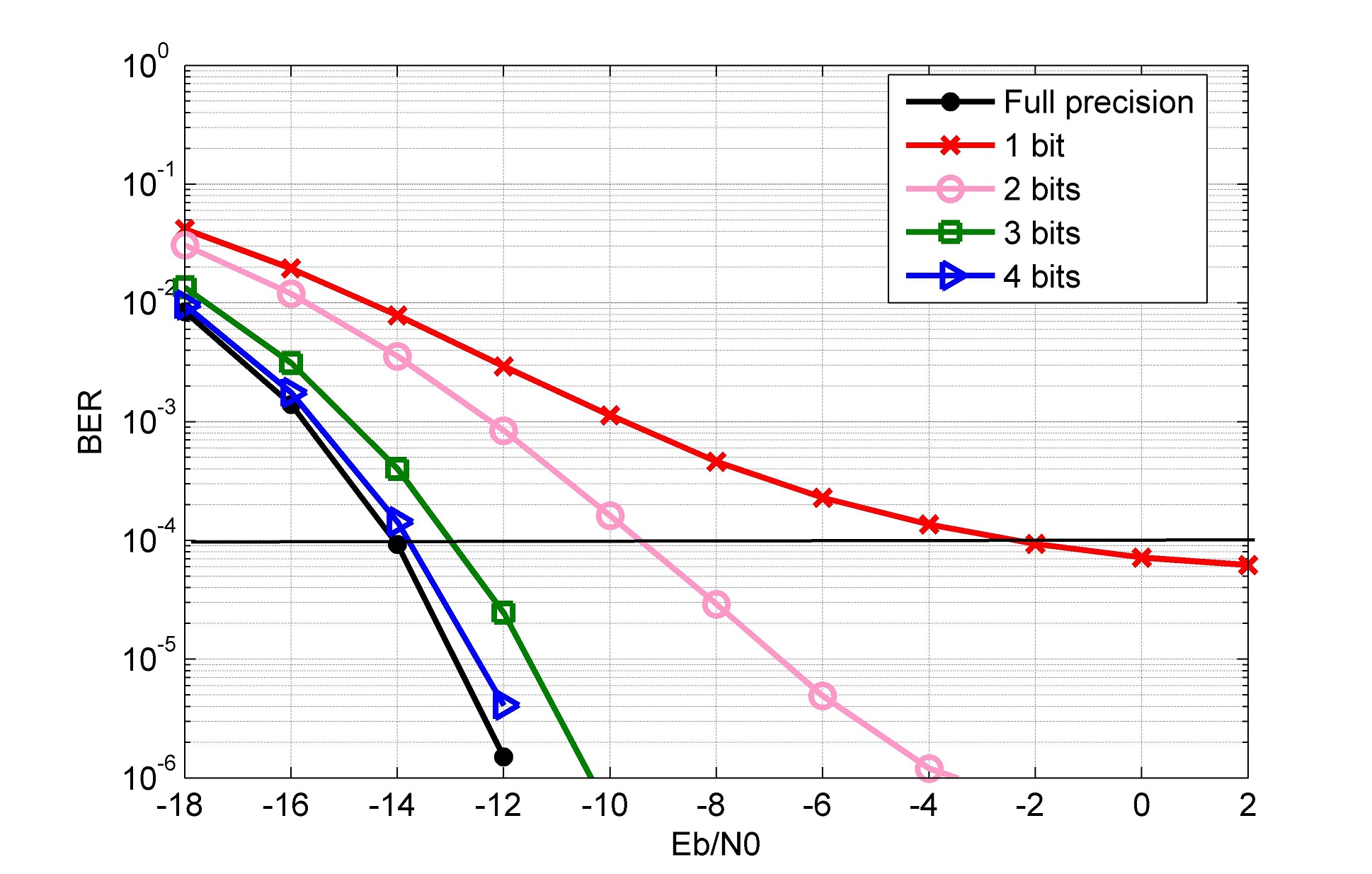}
\caption{BER versus $E_b/N_0$ using ZF detector for QPSK, $M=100$, $K=10$ and different ADC resolutions.}
\label{fig:ZF_QPSK}
\end{figure}

\begin{figure}[t!]
\centering
\includegraphics[width=0.99\linewidth]{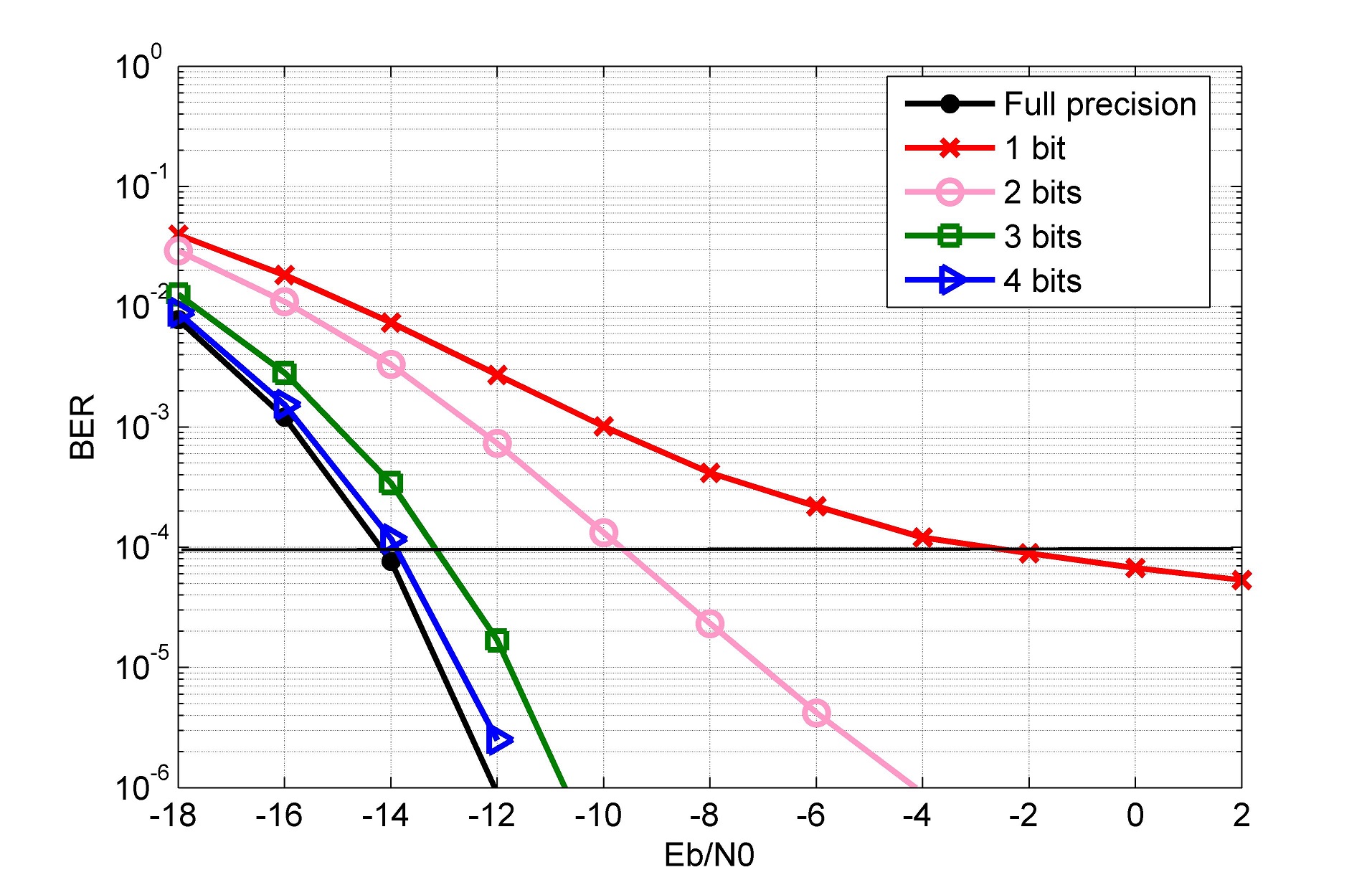}
\caption{BER versus $E_b/N_0$ using MMSE detector for QPSK, $M=100$, $K=10$ and different ADC resolutions.}
\label{fig:MMSE_QPSK}
\end{figure}

Assuming QPSK transmission modulation, the effect of quantization resolution on the BER performance versus the SNR per bit ($E_b/N_0$ in dB) for ZF and MMSE detection techniques are shown in Figures \ref{fig:ZF_QPSK} and \ref{fig:MMSE_QPSK}, respectively. As SNR grows, the BER decreases exponentially in a similar way for both cases of ZF and MMSE using the same ADC resolutions. However, for the lowest quantization resolution of 1-bit ADCs, even by increasing SNR the BER performance is limited to an approximate BER value of $10^{-4}$. We can see from theses figures that by increasing the resolution of quantizers, i.e 2-, 3- and 4-bit ADCs, the BER performance approaches the ideal case of having full precision ADCs.

\begin{figure}
\centering
\includegraphics[width=0.99\linewidth]{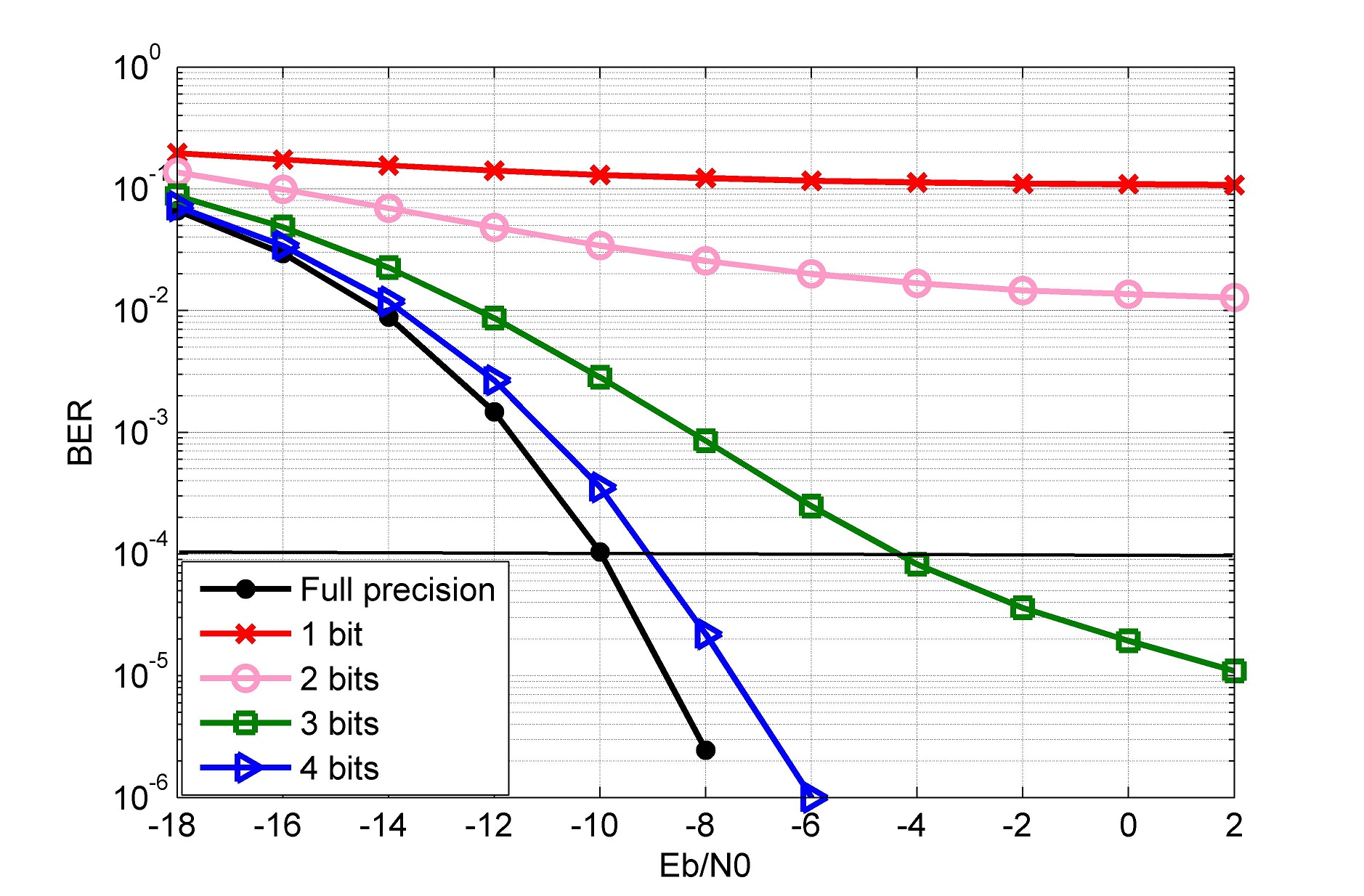}
\caption{BER versus $E_b/N_0$ using ZF detector for 16 QAM, $M=100$, $K=10$ and different ADC resolutions.}
\label{fig:ZF_16QAM}
\end{figure}

\begin{figure}
\centering
\includegraphics[width=0.99\linewidth]{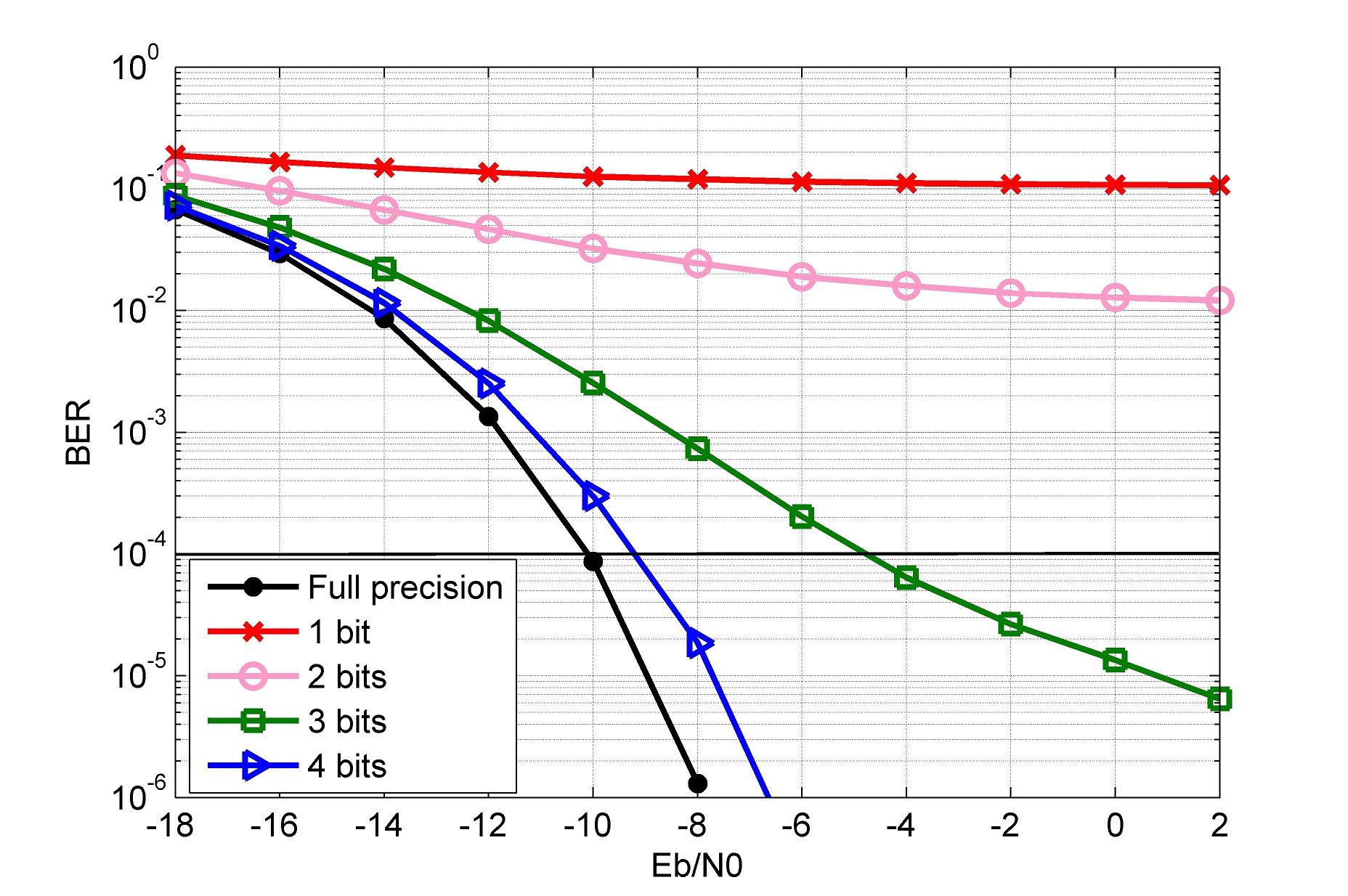}
\caption{BER versus $E_b/N_0$ using MMSE detector for 16 QAM, $M=100$, $K=10$ and different ADC resolutions.}
\label{fig:MMSE_16QAM}
\end{figure}

In order to investigate higher-order modulations, we plot the BER performance versus the SNR (in dB) for 16-QAM in the same way in Figure \ref{fig:ZF_16QAM} and Figure \ref{fig:MMSE_16QAM}. In this case, we observe that with 1-bit and 2-bit quantizations, the BER performance degrades severely. For the case of using 3-bit ADCs, somehow an acceptable BER performance is achieved. Although it has an increasing performance gap (compared to the ideal case) for larger values of SNR. However, having 4-bit quantization resolution and above, the BER performance is very close to the full precision quantization. Comparing the results for QPSK and 16-QAM, we see that when the modulation order increases, the BER increases accordingly for a specific ADC resolution. Therefore higher quantization resolution is required for higher order modulations to achieve the same performance.

\begin{figure}
\centering
\includegraphics[width=0.99\linewidth]{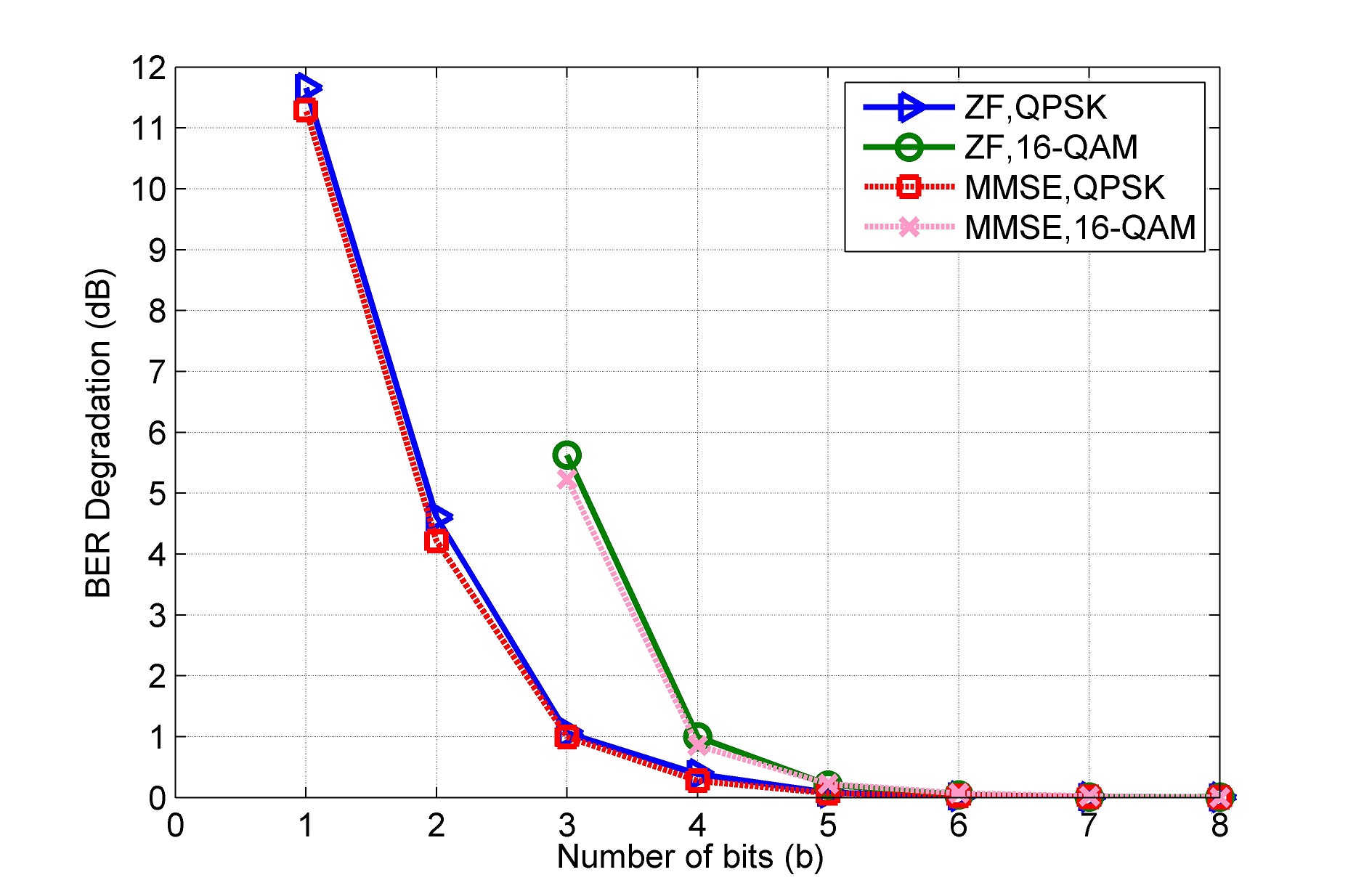}
\caption{BER degradation as a function of the ADC resolution for $M=100$ and $K=10$.}
\label{fig:BER_degrad_ADC_res}
\end{figure}

Fig. \ref{fig:BER_degrad_ADC_res} demonstrates the BER performance degradation as a function of the ADC resolution for both ZF and MMSE detectors. To define and plot the BER degradation, we first consider a reference SNR to achieve a BER value $10^{-4}$ for the full precision quantization. Then the extra SNR (in dB) required to get the same BER, for other ADC resolutions are calculated and depicted in the figure. We observe that using 1-bit ADCs in QPSK, a large extra SNR of about 11.5 dB is required to achieve the BER of $10^{-4}$, whereas using 2-bit ADCs in QPSK and 3-bit ADCs in 16-QAM require 4.5 to 5.5 dB SNR. Finally by employing higher resolution quantizers, over 3-bit ADCs for QPSK and 4-bit ADCs for 16-QAM, the BER degradation is around 1.5 dB or less.

\begin{figure}
\centering
\includegraphics[width=0.99\linewidth]{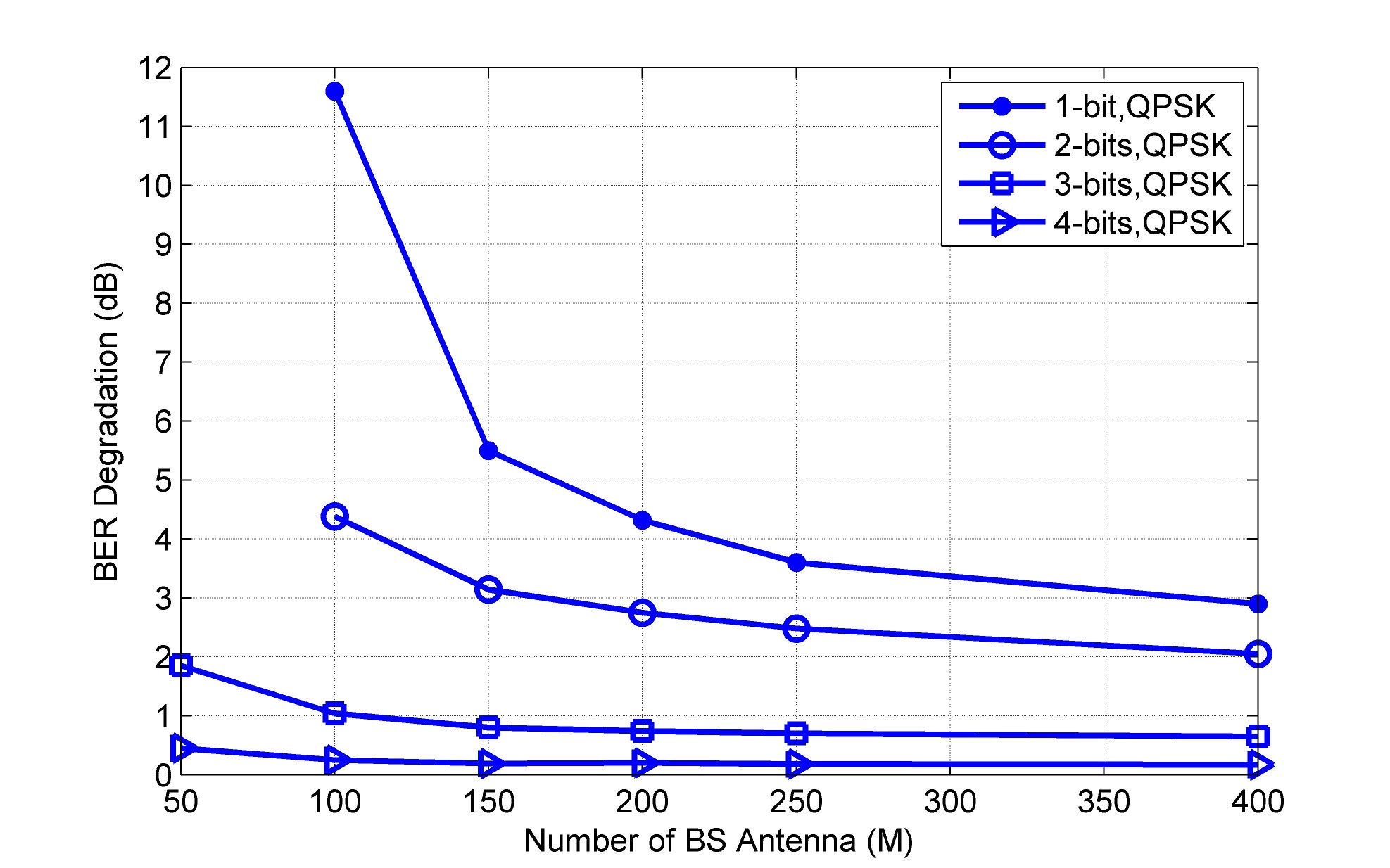}
\caption{BER degradation versus $M$ (number of BS antennas) using ZF detector for QPSK, and fixed $K=10$.}
\label{fig:BER_degrad_M_QPSK}
\end{figure}

\begin{figure}
\centering
\includegraphics[width=0.99\linewidth]{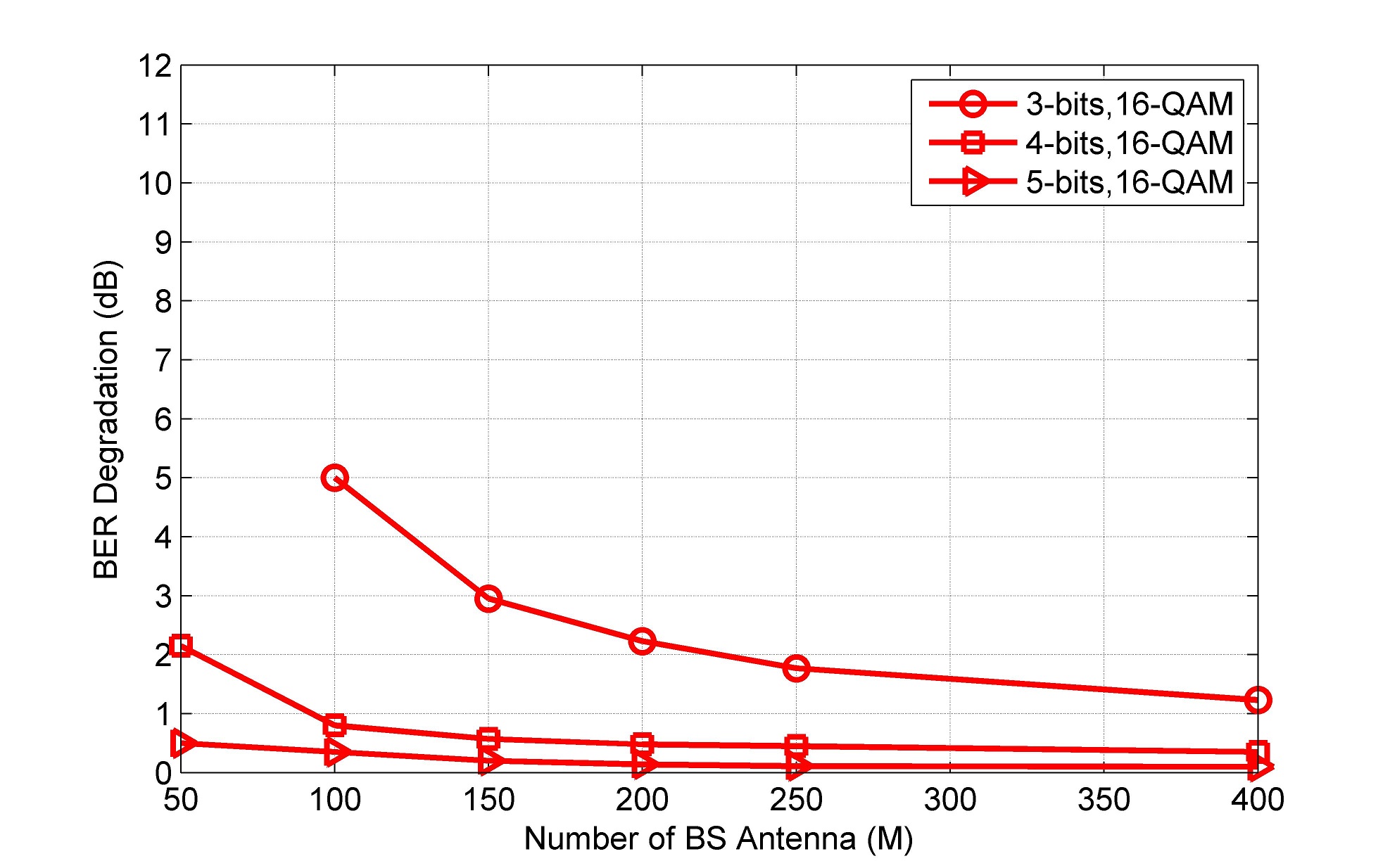}
\caption{BER degradation versus $M$ (number of BS antennas) using ZF detector for 16-QAM, and fixed $K=10$.}
\label{fig:BER_degrad_M_16QAM}
\end{figure}

Another simulation is performed to determine the BER degradation as a function of ADC resolution and the number of BS antenna, and results are illustrated in Figures \ref{fig:BER_degrad_M_QPSK} and \ref{fig:BER_degrad_M_16QAM}. Due to the negligible differences in the performance of ZF and MMSE detectors, we only employ ZF detection in these two figures. We consider a fixed number of users, $K$ = 10, but M is varied from 50 to 400. It can be seen that, by increasing the number of BS antennas, the BER degradation caused by low-resolution ADCs can be reduced. 
For example, in Fig. \ref{fig:BER_degrad_M_QPSK} a 3-bit quantized system with $M=50$ has the same BER performance as the case of 2-bit quantized with a roughly $M=400$ number of antennas at the BS. Similarly, we see that in Fig. \ref{fig:BER_degrad_M_16QAM}, the BER performance of a 4-bit quantized system with $M=50$ is approximately achieved by a 3-bits ADCs employing $M=200$ BS antennas. Therefore, higher M is suggested to compensate the BER performance degradation in the case of having fixed quantization levels of low-resolution ADCs. However, this increase in M depends on both the modulation type and the number of bits in our quantization.

\section{Conclusion}\label{sec:conc}
In this paper, we investigated the effect of low resolution ADCs on the BER performance of uplink massive MIMO systems by carrying out simulations, for the two modulation types of QPSK and 16-QAM. Our results revealed that massive MIMO technology with low resolution $b$-bit ADCs ($b$ from 1 to 4 bits) can achieve acceptable performances compared to the case of using high resolution ADCs. 
We also realized that having more antennas at BS cancels out the performance reduction effect of using low-resolution ADCs on BER. 
However, this increase in M depends on both the modulation type and the number of bits in our quantization. Therefore, low resolution ADCs can be employed in massive MIMO systems as a solution to reduce the cost and the power consumption of the BS.


\bibliographystyle{IEEEtran}

\end{document}